\documentclass[twocolumn]{aastex7}

\usepackage{amsmath}
\usepackage{multirow}
\usepackage{threeparttable}
\usepackage{CJK}

\newcommand{\oiii}{[O\,{\sc iii}]}

\newcommand{\ha}{H$\alpha$}

\newcommand{\sii}{[S\,{\sc ii}]}

\begin{document}
\begin{CJK*}{UTF8}{gbsn}
\title{Local Analogs of Little Red Dots: Optical Variability and Evidence for an Active Galatic Nucleus Origin}

\author[0000-0003-3987-0858]{Ruqiu Lin ({\CJKfamily{gbsn}林如秋})}\email[show]{ruqiulin@umass.edu}
\affiliation{Shanghai Astronomical Observatory, Chinese Academy of Sciences, 80 Nandan Road, Shanghai 200030, People’s Republic of China}
\affiliation{University of Massachusetts Amherst, 710 North Pleasant Street, Amherst, MA 01003-9305, USA}

\author[0000-0002-9634-2923]{Zhen-Ya Zheng*}
\affiliation{Shanghai Astronomical Observatory, Chinese Academy of Sciences, 80 Nandan Road, Shanghai 200030, People’s Republic of China}
\email[show]{email: zhengzy@shao.ac.cn}
\correspondingauthor{Zhen-Ya Zheng, Ruqiu Lin}

\author[0000-0002-4419-6434]{Junxian Wang}\email{jxw@ustc.edu.cn}
\affiliation{Department of Astronomy, University of Science and Technology of China, Hefei, Anhui 230026, China}

\author[0000-0001-6947-5846]{Luis C. Ho}\email{lho.pku@gmail.com}
\affiliation{Kavli Institute for Astronomy and Astrophysics, Peking University, Beijing 100871, China}
\affiliation{Department of Astronomy, School of Physics, Peking University, Beijing 100871, China}

\author[0000-0002-7051-1100]{Jorge A. Zavala}\email{jzavala@umass.edu}
\affiliation{University of Massachusetts Amherst, 710 North Pleasant Street, Amherst, MA 01003-9305, USA}

\author[0000-0002-2420-5022]{Zijian Zhang}\email{zjz.kiaa@stu.pku.edu.cn}
\affiliation{Kavli Institute for Astronomy and Astrophysics, Peking University, Beijing 100871, China}
\affiliation{Department of Astronomy, School of Physics, Peking University, Beijing 100871, China}

\author[0000-0002-0003-8557]{Chunyan Jiang}\email{cyjiang@shao.ac.cn}
\affiliation{Shanghai Astronomical Observatory, Chinese Academy of Sciences, 80 Nandan Road, Shanghai 200030, People’s Republic of China}

\author[0000-0002-3134-9526]{Jiaqi Lin}\email{linjiaqi@shao.ac.cn}
\affiliation{School of Physics and Astronomy, Sun Yat-sen University, Zhuhai 519082, People’s Republic of China}
\affiliation{Shanghai Astronomical Observatory, Chinese Academy of Sciences, 80 Nandan Road, Shanghai 200030, People’s Republic of China}
\affiliation{School of Astronomy and Space Sciences, University of Chinese Academy of Sciences, No. 19A Yuquan Road, Beijing 100049, People's Republic of China}

\author[0000-0001-6763-5869]{Fang-Ting Yuan}\email{yuanft@shao.ac.cn}
\affiliation{Shanghai Astronomical Observatory, Chinese Academy of Sciences, 80 Nandan Road, Shanghai 200030, People’s Republic of China}

\author[0000-0003-4176-6486]{Linhua Jiang}\email{jiangKIAA@pku.edu.cn}
\affiliation{Kavli Institute for Astronomy and Astrophysics, Peking University, Beijing 100871, China}
\affiliation{Department of Astronomy, School of Physics, Peking University, Beijing 100871, China}

\author[0000-0002-1517-6792]{Tinggui Wang}\email{twang@ustc.edu.cn}
\affiliation{Department of Astronomy, University of Science and Technology of China, Hefei, Anhui 230026, China}

\author[0000-0002-1542-8080]{Xiaer Zhang}\email{zxe@shao.ac.cn}
\affiliation{Shanghai Astronomical Observatory, Chinese Academy of Sciences, 80 Nandan Road, Shanghai 200030, People’s Republic of China}

\begin{abstract}
Little red dots (LRDs) draw extensive attention because of their unique observational characteristics and apparent overabundance in the early Universe, raising new insights into early black hole formation and growth. Early studies show that LRDs exhibit weak variability in photometry and emission-line fluxes, suggesting a preference for super-Eddington accretion or disfavouring an active galatic nucleus (AGN) origin. However, the current data is limited, preventing us from placing strong constraints on their variability.
Based on Zwicky Transient Facility light curves with a baseline of $\sim6$ yr, we here study the optical variability of seven previously reported local analogs of LRDs at $z \sim 0.3$, offering an insight into LRDs from a low-redshift sample. 
Three out of seven local analogs show excess variances on all three bands of their light curves, and two of them can be fitted with the damping random walk model, supporting their AGN origins for the variability. The remaining sources show weak variance in at least one band, but no detectable variability, exhibiting $\rm SF_\infty$ upper limits consistent with estimates from high-redshift LRDs. Their nondetection of variability is likely due to the large photometric uncertainty.
As an implication, by simulating long baseline light curves with the variability amplitude of local analogs and adopting JWST observation cadence, we investigate the limitation of the variability amplitude estimate for LRDs. Our mock observations imply that the current constraints on LRDs' variability are probably underestimated. This underestimation might be induced by the short temporal baseline of observations, as well as the intrinsic scatter of the empirical $M_{\rm BH}-\tau$ relation.
\end{abstract}



\section{Introduction}
The superb sensitivity of the James Webb Space Telescope (JWST) has allowed the identification of the so-called ``little red dots'' \citep[LRDs,][]{Matthee2024ApJ...963..129M}, one of the most mysterious objects in the high-redshift (high-$z$) Universe. LRDs are identified with extremely compact morphologies and ``V-shaped'' UV-optical spectral energy distributions  \citep[SEDs][]{Labbe2025ApJ...978...92L, Akins2023ApJ...956...61A, Barro2024ApJ...963..128B}. They have little evidence for an extended host-galaxy component \citep{Chen2025ApJ...983...60C, Zhang2025arXiv251025830Z}. In addition, the origin of their relatively blue ultraviolet SEDs remains debated \citep[see review in][]{Inayoshi2025arXiv251203130I}. Follow-up spectroscopic studies have revealed a high incidence of broad emission lines, unusually large viral black hole-to-host mass ratios ($M_{\rm BH}/M_{*}$) \citep{Kocevski2023ApJ...954L...4K, Matthee2024ApJ...963..129M, Harikane2023ApJ...959...39H,  Maiolino2024A&A...691A.145M, Pacucci2023ApJ...957L...3P}, and, in some cases, prominent absorption Balmer features \citep{Matthee2024ApJ...963..129M}. These properties distinguish LRDs from classical type I AGNs and suggest that they may represent a distinct or transitional population.

Since the discovery of LRDs, their radiation origin, whether their emission is primarily driven by AGNs, has been the subject of intense debate. On the one hand, many groups have continued to expand LRD samples, extending their redshift coverage down to $z \sim 2$ \citep{Euclid2025arXiv250315323E, Kocevski2025ApJ...986..126K, Ma2025arXiv250408032M}, and have investigated their evolutionary properties through both theoretical modeling and observational constraints \citep{Inayoshi2025ApJ...980L..27I, Li2025ApJ...980...36L, Naidu2025arXiv250316596N, Wang2025arXiv251109278W, Zhang2025arXiv250512719Z}. On the other hand, numerous studies have focused on their multiwavelength characteristics in an effort to independently probe the physical nature of LRDs and assess whether they can be unambiguously classified as AGNs \citep{Yue2024ApJ...974L..26Y, Perez-Gonzalez2024ApJ...968....4P, Casey2024ApJ...975L...4C, Akins2025ApJ...980L..29A, Casey2025ApJ...990L..61C}.
Using image stacking techniques, studies have reported weak X-ray and radio emission from LRDs, which has raised concerns regarding their AGN origin or hints of an obscured, Compton-thick AGN with super-Eddington accretion \citep{Yue2024ApJ...974L..26Y, Perger2025A&A...693L...2P, Gloudemans2025ApJ...986..130G}. 

More recently, based on multi-epoch observations obtained with JWST over a two-year baseline, \cite{Kokubo2025ApJ...995...24K} derived upper limits on the rest-frame optical variability amplitudes of a small sample of LRDs and concluded that their variability is weak. Based on methodology similar to that of \cite{Kokubo2025ApJ...995...24K}, \cite{Zhang2025ApJ...985..119Z} extended the LRD sample and revealed the significant variability of a handful of LRDs.
In contrast, \cite{Tee2025ApJ...983L..26T} revealed weak rest-frame UV variability of a sample of LRDs. Additionally, exploiting gravitational lensing, \cite{Furtak2025A&A...698A.227F} monitored the broad-line fluxes of individual LRDs through lensed multiple images and detected variability with a rest-frame timescale of $\sim 2.4$ yr. Notably, \cite{Zhang2025arXiv251205180Z} revealed year-scale variability in a lensed LRD.
Although stochastic variability is widely regarded as a robust signature of AGN activity, the currently available observations are limited by short temporal baselines or are restricted to one or two objects. 

To better understand the nature of LRDs, efforts have focused on the search for local analogs (hereafter ``{\it local LRDs}'' for convenience) in the nearby Universe and have attempted to investigate their physical properties across multiple wavelengths \citep{Lin2025ApJ...980L..34L, Lin2025arXiv250710659L, Ji2025MNRAS.tmp.2102J, Chen2025arXiv251002801C}. \cite{Lin2025ApJ...980L..34L} established a direct connection between LRDs and low-redshift strong emission-line galaxies, identifying seven local AGNs. These sources exhibit striking similarities to high-$z$ LRDs in terms of broad-line widths, SED shapes, black hole masses, ultraviolet luminosities, and accretion rates, thereby providing a valuable reference sample for detailed studies of LRD properties.
Subsequently, \cite{Lin2025arXiv250710659L} and \cite{Ji2025MNRAS.tmp.2102J} independently selected additional local LRDs from previously known samples of low-redshift metal-poor AGNs. \cite{Chen2025arXiv251002801C} also discovered a local analog from a sample of ultraluminous infrared galaxies. Follow-up high-sensitivity spectroscopy with ground-based telescopes revealed that some local LRDs also appear to have Balmer absorption features \citep{Lin2025arXiv250710659L}.
However, local LRDs face challenges similar to those of high-$z$ LRDs, particularly regarding the ionization sources of their UV emission and broad lines (AGN $vs.$ star formation), as well as the physical nature underlying multiband properties. 

Variability provides a chance to distinguish AGN from star formation. Most known local LRDs have been identified from large ground-based spectroscopic surveys and are generally covered by time-domain surveys such as Zwicky Transient Facility (ZTF), providing a better opportunity to unveil their variability. \cite{Lin2025arXiv250710659L} reviewed optical and infrared light curves of three local LRDs and found no significant variability.
However, there is still a lack of systematic investigation of the variability properties of local LRDs.

To better characterize the variability properties of local LRDs and to provide a meaningful reference for interpreting variability constraints, in this work, we present a dedicated variability analysis based on the local analogs of LRDs from \cite{Lin2025ApJ...980L..34L}. We describe the sample and light-curve data in Section \ref{sec:sampledata}. The method to estimate the variability and light-curve modeling are presented in Section~\ref{sec:methods}. In Section~\ref{sec:AGNorigin}, we reveal the AGN origin of local LRDs suggested by our analyses. We claim the current underestimation of variability for high-$z$ LRDs through simulating light curves in Section~\ref{sec:underestimation}. We conclude this work in Section~\ref{sec:conclusion}.

\section{Sample and Data}\label{sec:sampledata}
\subsection{Low-mass AGNs and Local LRDs in Green Pea Galaxies}
We start from a massive black hole (MBH) sample constructed in \cite{Lin2024SCPMA..6709811L}, which contains Green Pea galaxies (GPs) with broad \ha\ emission at $z<0.4$ selected from Large Sky Area Multi-Object Fiber Spectroscopic Telescope (LAMOST) and Sloan Digital Sky Survey (SDSS) spectroscopic surveys. To briefly summarize the selection method, we first subtracted the continuum emission using {\tt pPXF} based on composite spectrum models. We then decomposed their \ha\ emission lines into narrow, wide, and broad (if present) components, referenced by the narrow-line profiles of \sii\ or \oiii. From 2190 GP spectra, we identified 59 candidate MBHs with broad \ha\ emission. For these sources, we adopted the parameter measurements (e.g., Black hole mass) derived in \cite{Lin2024SCPMA..6709811L}.

\begin{figure*}
    \centering
    \hspace{-1cm}
    \includegraphics[width=0.95\linewidth]{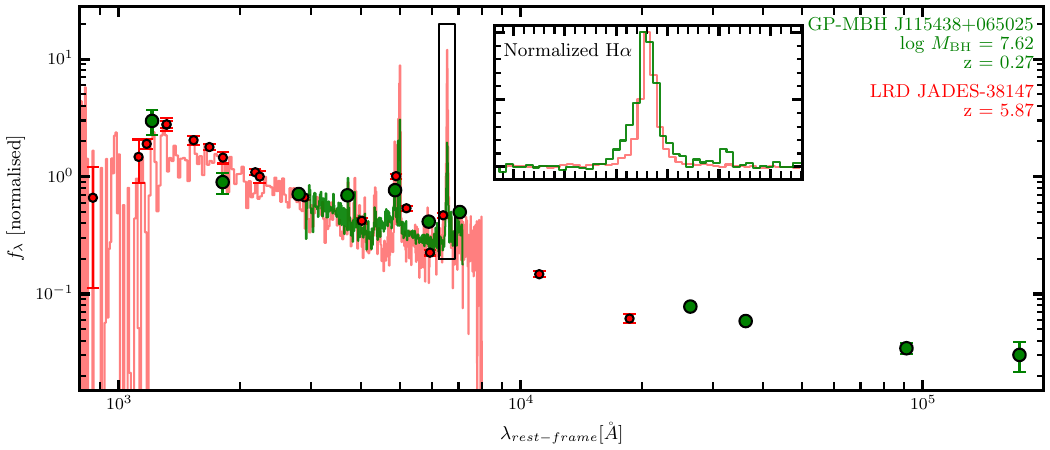}
    \caption{An example of the SED of the local analog to LRDs. Green line and dots are the SDSS spectrum and UV-IR photometries of J115438+065025. Red line and dots are JWST/NIRSpec prism spectrum and NIRCam and MIRI photometry of an LRD JADES-38147 \citep{Setton2025ApJ...995..118S}, retrieved from the Dawn JWST Archive (DJA\footnote{https://dawn-cph.github.io/dja/index.html}, \citealt{Valentino2023ApJ...947...20V, de_Graaff2025A&A...697A.189D, Heintz2024Sci...384..890H}). The SDSS spectrum is rebinned to meet the spectral resolution of JADES-38147's spectrum. The zoom-in window shows the broad-line \ha\ of these two sources.}
    \label{fig:spec}
\end{figure*}

Subsequently, \cite{Lin2025ApJ...980L..34L} identified seven local analogs of LRDs within this sample, based on similar observational characteristics, i.e., V-shaped UV-optical SED, broad-line emission, and compact morphology. We use the term ``local LRDs'' to refer to local analogs of LRDs for convenience. For a detailed discussion of the uncertainties in the sample selection, we refer the reader to \citet{Lin2025ApJ...980L..34L}. In Figure~\ref{fig:spec} we show an example of their SEDs covering rest-frame wavelength from $\sim 0.1$ to $20\, \mu m$. For comparison, we also show the best-match LRD's SED taken in the JWST Advanced Deep Extragalactic Survey (JADES, PI: Eisenstein, GTO\#1181) in red, which was reported in \cite{Setton2025ApJ...995..118S}. 
            
\begin{figure*}
    \centering
    \includegraphics[width=0.48\linewidth]{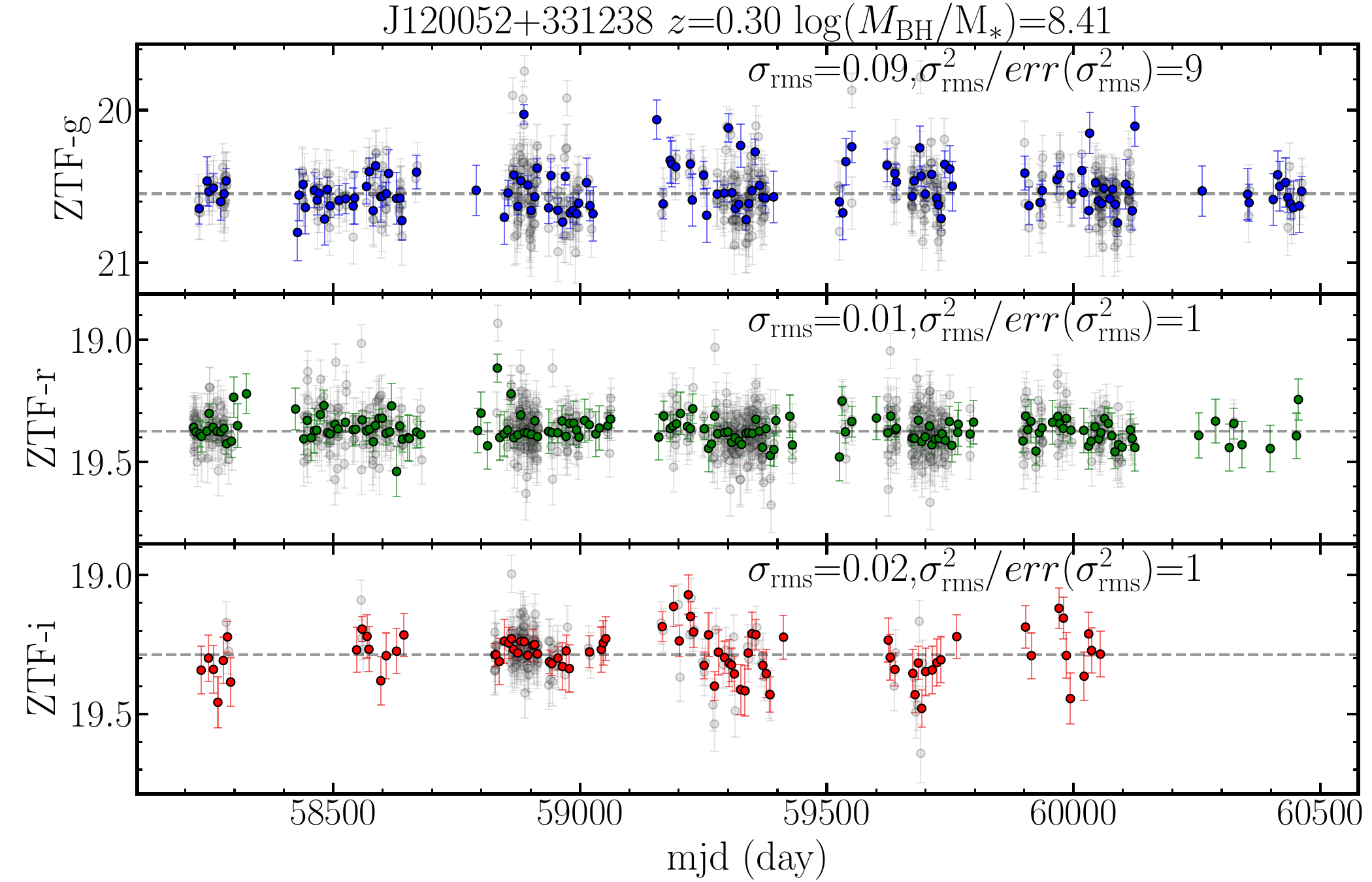}
    \includegraphics[width=0.48\linewidth]{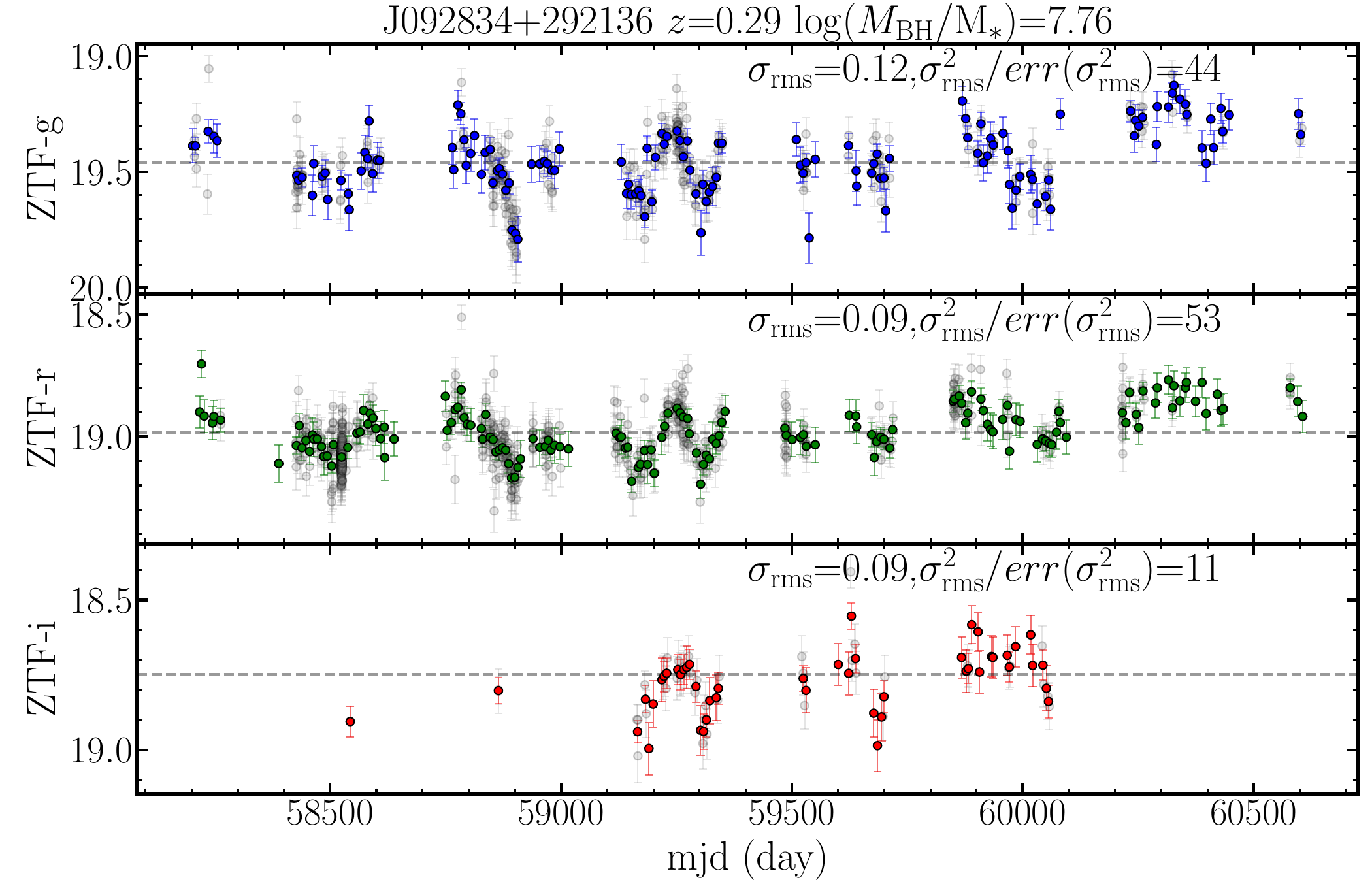}
    \includegraphics[width=0.48\linewidth]{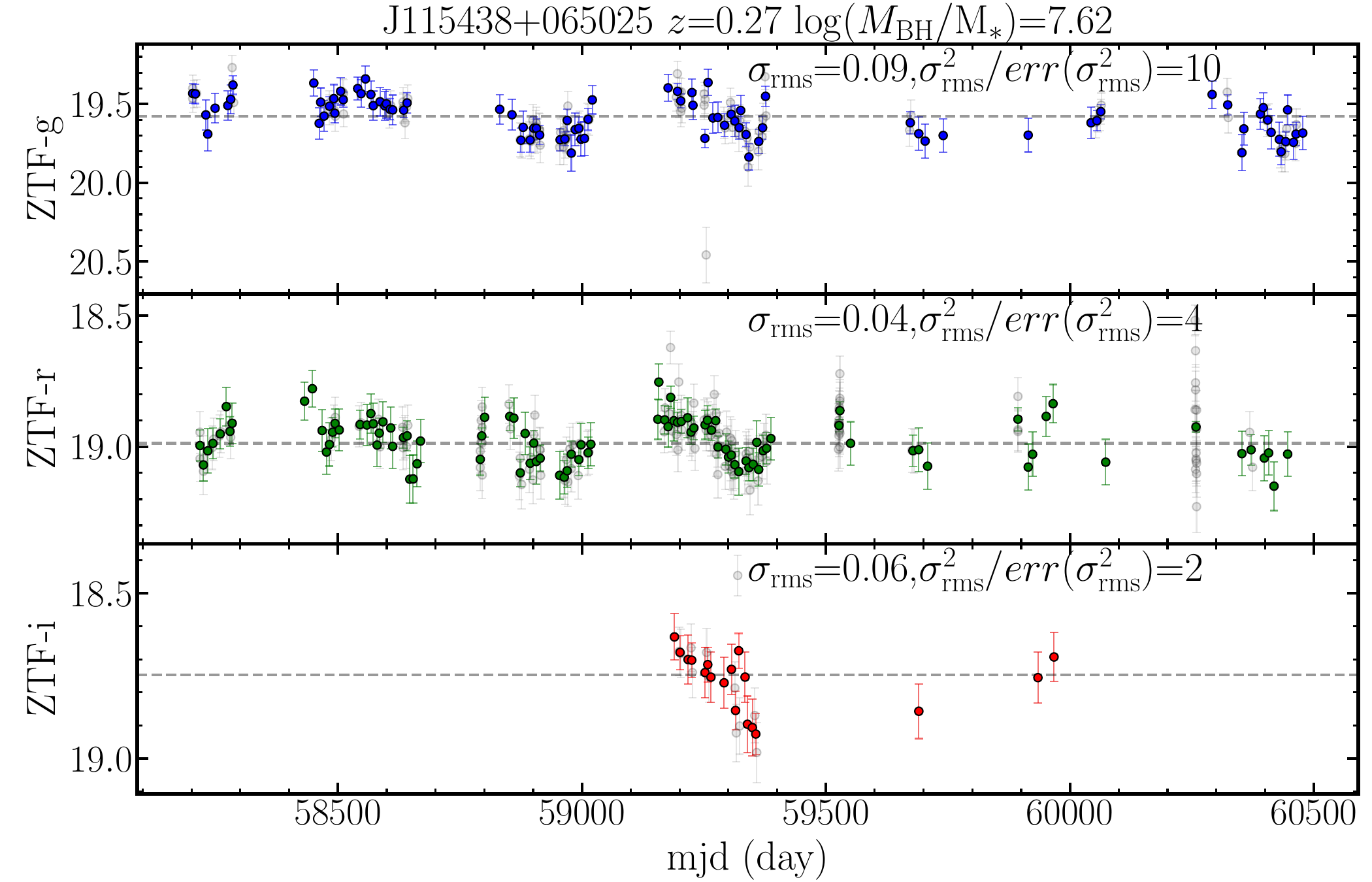}
    \includegraphics[width=0.48\linewidth]{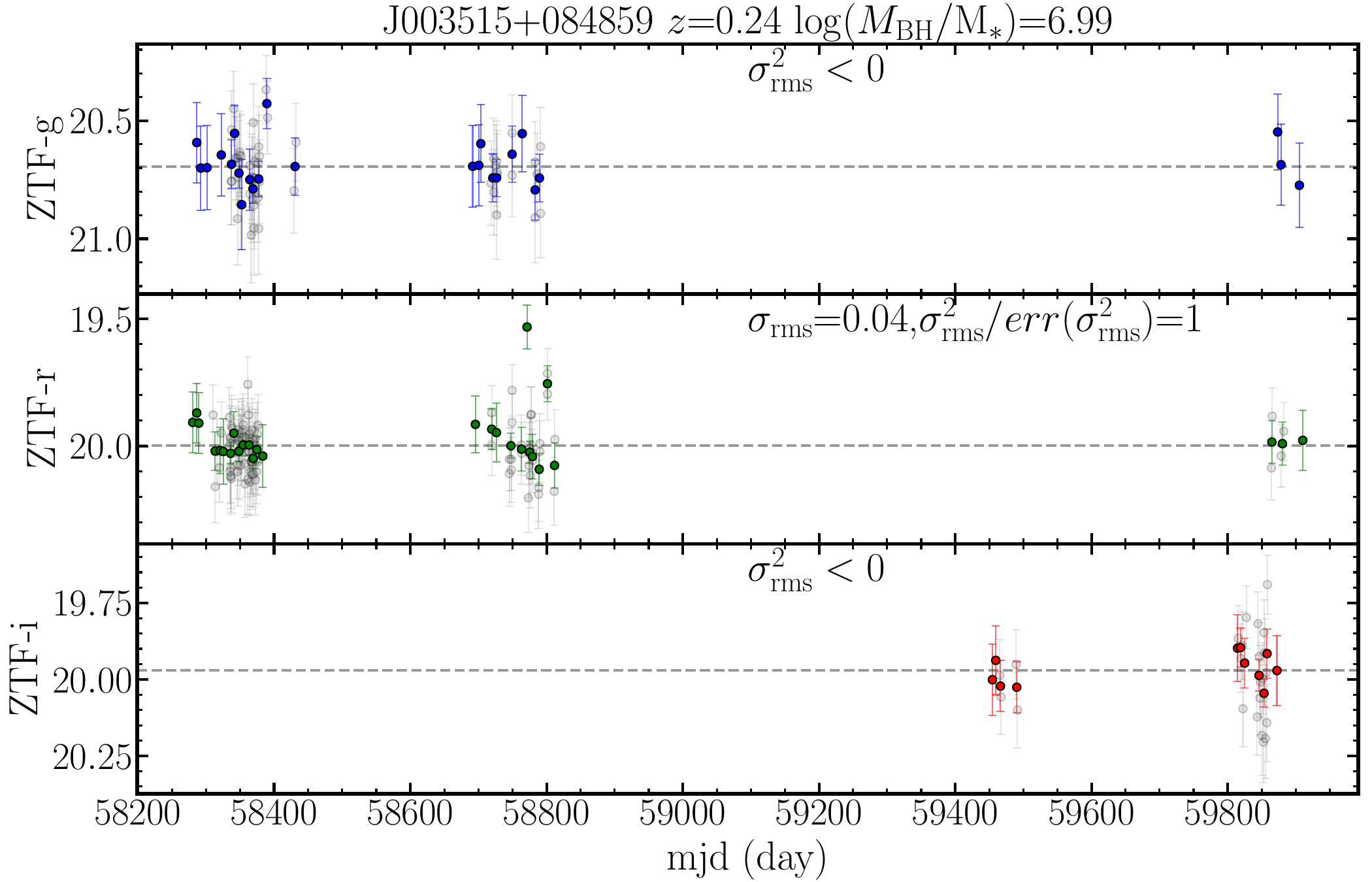}
    \includegraphics[width=0.48\linewidth]{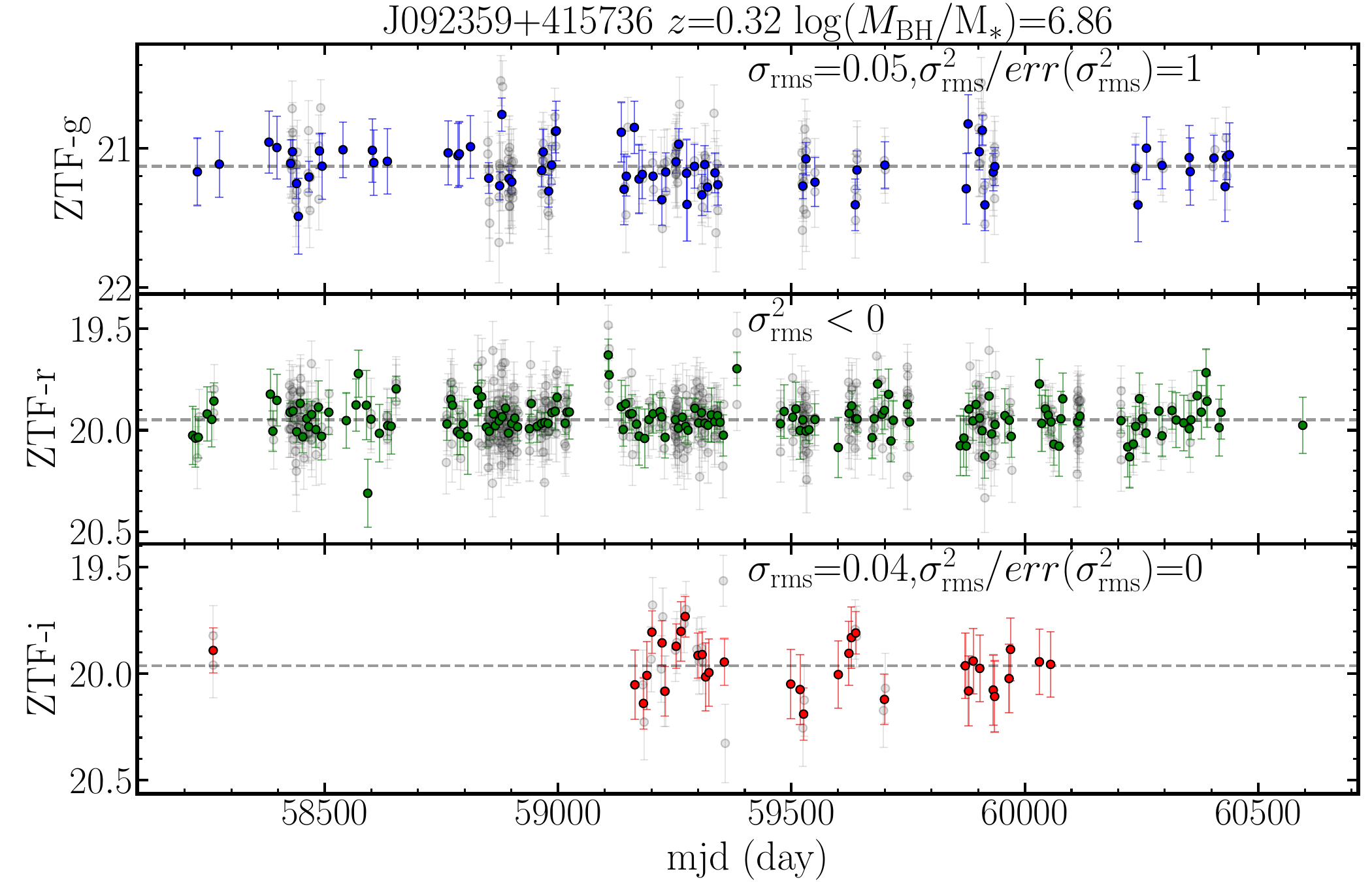}
    \includegraphics[width=0.48\linewidth]{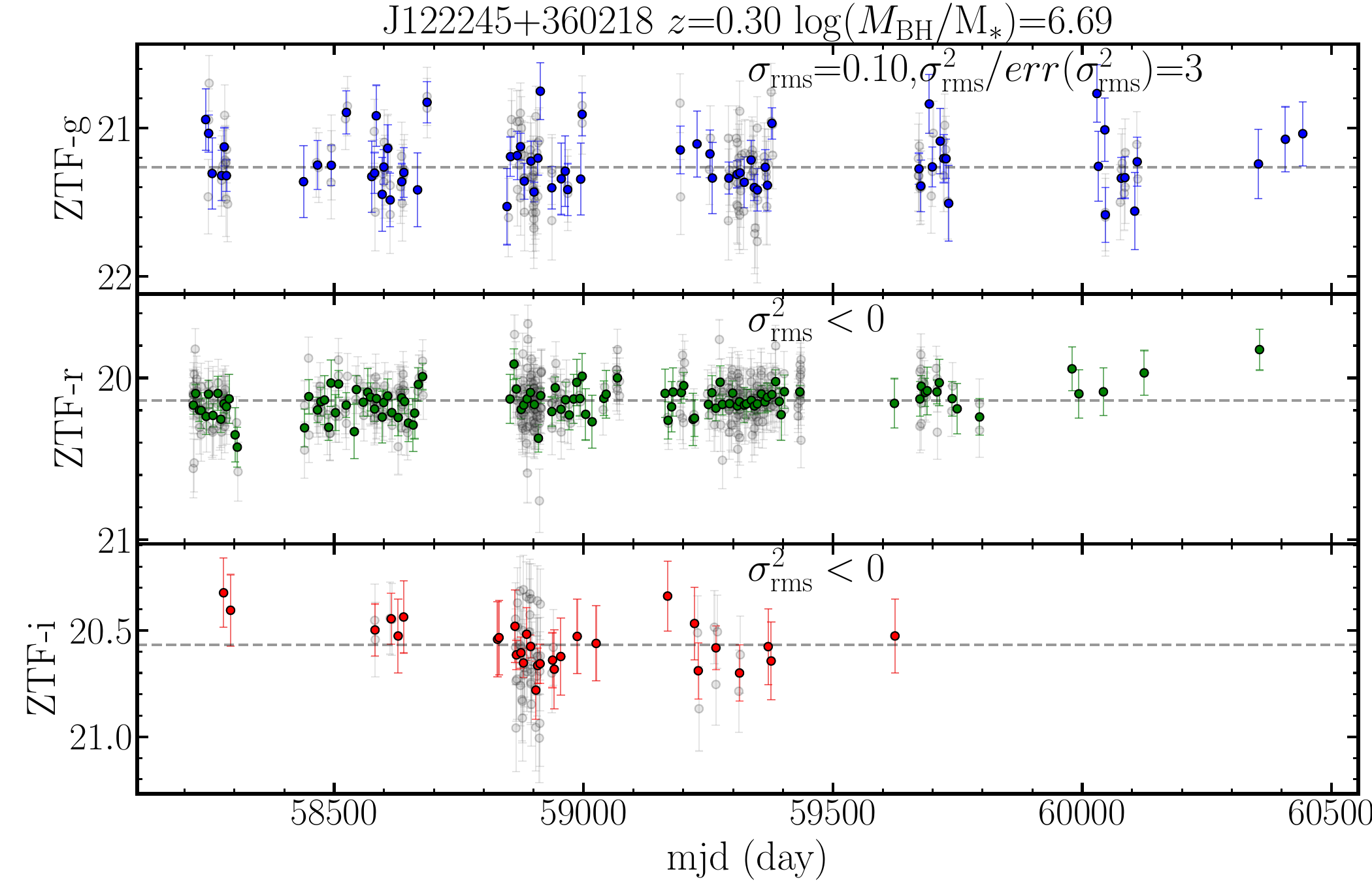}
    \includegraphics[width=0.48\linewidth]{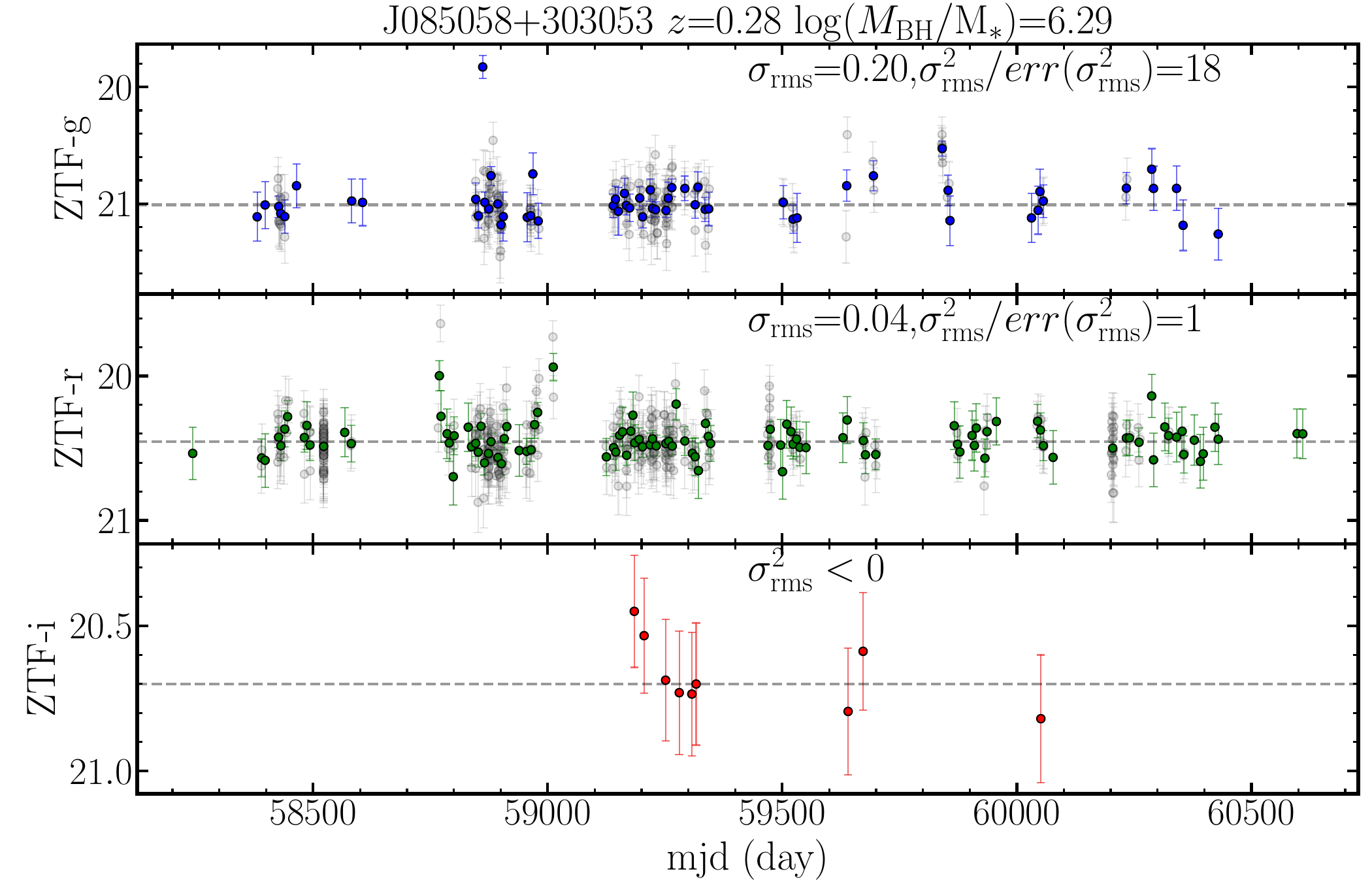}
    \caption{ZTF light curves of local LRDs. Gray dots are measurements after removing data with {\tt catflags} $\neq 0$ and {\tt airmass} $>2$, and colored dots are rebinned by 7 days, while colors blue, green, and red correspond to $g$, $r$, and $i$ bands, respectively. The weighted excess variance $\sigma_{\rm rms}$ and $\sigma_{\rm rms}^2$/err($\sigma_{\rm rms}^2$) for each band are indicated in the upper right corner.}
    \label{fig:lc}
\end{figure*}

\subsection{ZTF Light Curves} \label{sec:lc}
We adopt the light curves from ZTF and require the flag \texttt{catflags} $=0$. As those measurements with a large \texttt{airmass} have significant scatter, we remove those having an \texttt{airmass} $>2$. After that, most (six out of seven) have the number of measurements over 150, spanning $> 5$ yr, which can constrain variability given the current photometric error. The ZTF light curves have nonuniform temporal sampling due to the observing altitudes from the ground. Our sources have a highest cadence of less than a day, an observing window of 100-200 days, and an observing gap of 100 days. These observational effects may induce extra features in the correlated function of their light curves, such as pits at a timescale of $\sim 200$ days.

ZTF uses the PSF-corrected method to measure the photometry and to construct light curves. Based on the Pan-STARRS source catalog, they selected several reference stars for flux calibration. ZTF's single imaging have $5\sigma$ detection limits of $21.067\pm0.003$, $21.012\pm0.002$, and $20.51\pm0.01$ mag in $g$, $r$, and $i$ bands, respectively \citep{Masci2019PASP..131a8003M}. Figure~\ref{fig:lc} shows the light curves of all sources. The median $g$-band photometry are 20.56, 19.47, 19.59, 20.69, 21.15, 21.28, and 21.03 mag for J120052+331238, J092834+292136, J115438+065025, J003515+084859, J092359+415736, J122245+360218, and J085058+303053, respectively. This indicates that these sources are faint in ZTF data.

\section{Methods}\label{sec:methods}
In this section, we analyze the light curves using various methods, including weighted excess variance, damping random walk (DRW) model fitting, and upper-limit variability estimate. 

\subsection{Weighted Excess Variance}
To evaluate the variance of light curves without assuming any model, we follow \cite{Kang2024ApJ...971...60K} to estimate weighted excess variance $\sigma_{\rm rms}$ and corresponding uncertainty err($\sigma_{\rm rms}$). Given the faintness of these sources (see \ref{sec:lc} for details), we bin light curves by 7 days to increase the signal-to-noise ratio (S/N). We then follow the methodology of \cite{Kang2024ApJ...971...60K} (Appendix A) and adopt $w=-4$. The estimated $\sigma_{\rm rms}$ and S/N of $\sigma_{\rm rms}^2$, i.e., $\sigma_{\rm rms}^2$/err($\sigma_{\rm rms}^2$) for each band are indicated in Figure~\ref{fig:lc}.

\subsection{DRW Model Fitting}\label{sec:drwmodel}
The stochastic variability of AGNs is widely analyzed with a continuous autoregressive moving-average (CARMA) model. We use the simplest CARMA model, i.e., the DRW model, to describe AGN-like light curves. In this sample, only J092834+292136 and J115438+065025 show $\sigma_{\rm rms}^2$/err($\sigma_{\rm rms}^2$) $\geq 2$ in all three bands, indicating significant excess variance beyond photometric uncertainties. We fit their light curves with the DRW model using \texttt{celerite} \citep{celerite1, celerite2} and estimate the uncertainty using \texttt{emcee} \citep{Foreman-Mackey2013PASP..125..306F}. 

\begin{figure*}
    \centering
    \includegraphics[width=0.6125\linewidth]{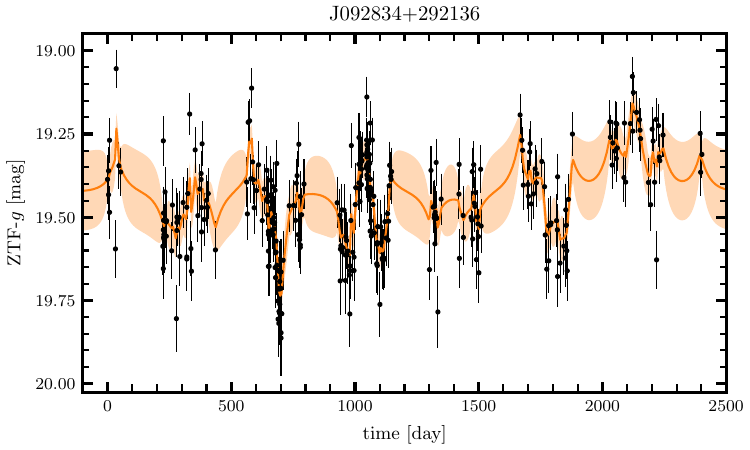}
    \includegraphics[width=0.3675\linewidth]{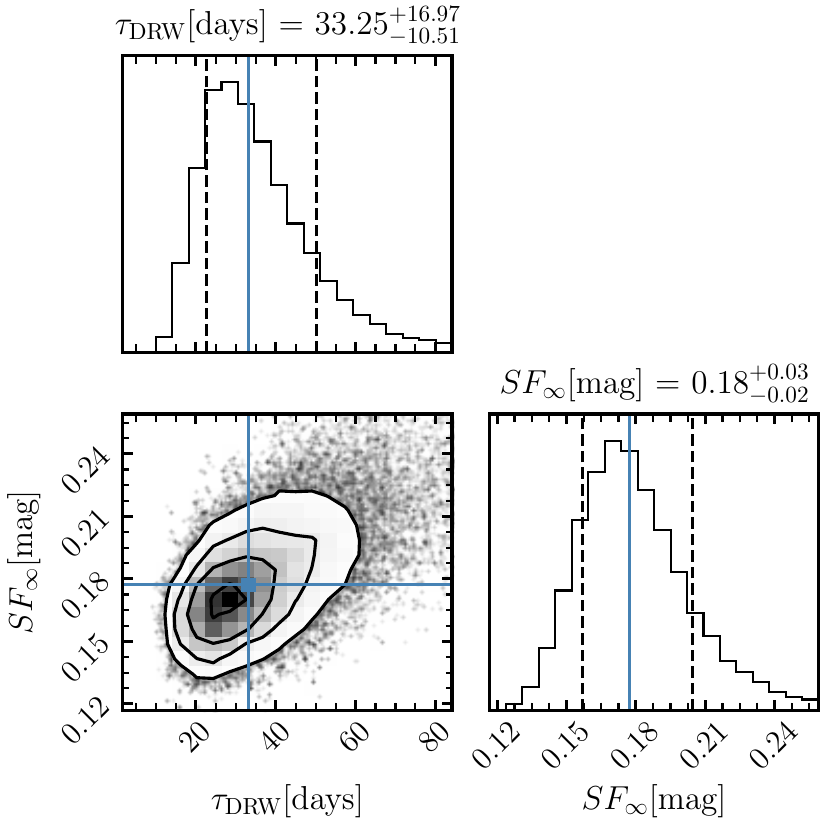}
    \includegraphics[width=0.6125\linewidth]{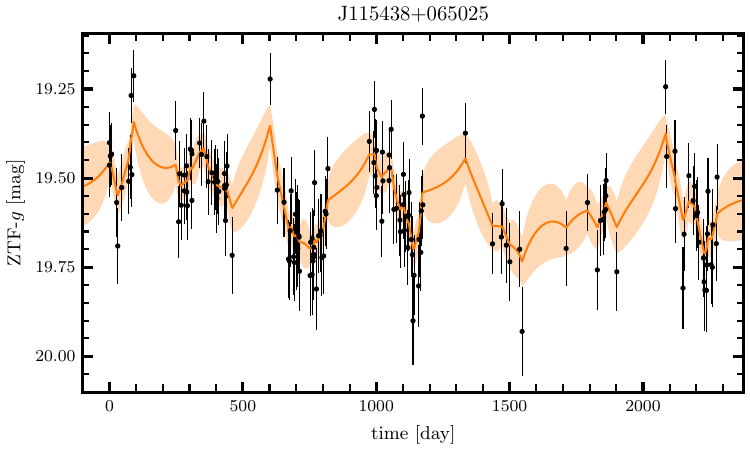}
    \includegraphics[width=0.3675\linewidth]{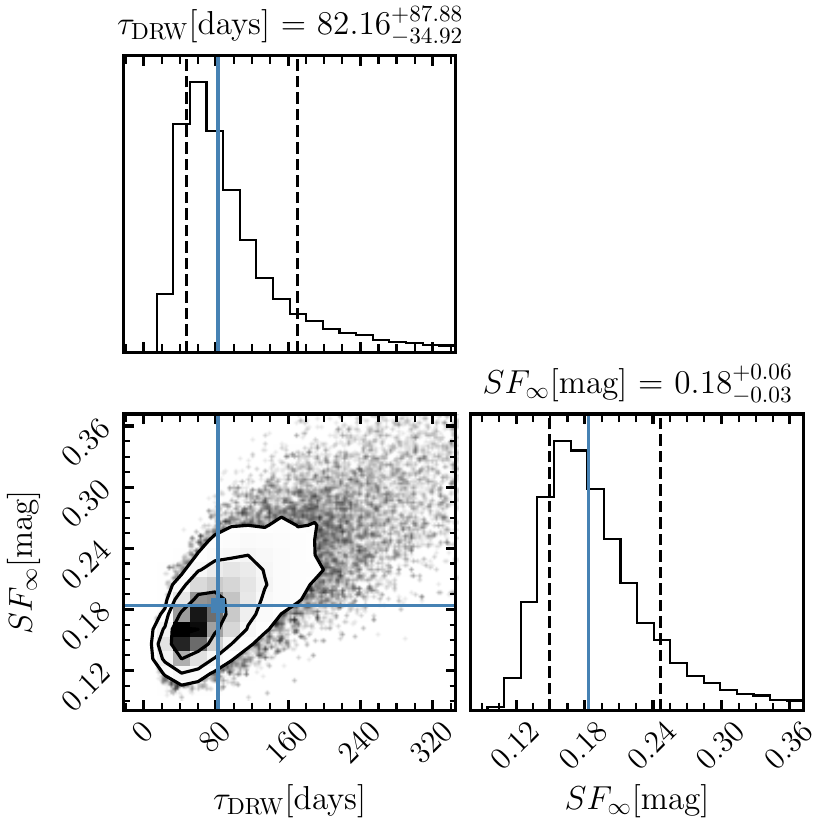}
    \caption{Light-curve modeling of J092834+292136 ({\it top}) and J115438+065025 ({\it bottom}). The left panels show the observed light curves (black dots) together with the best-fitting DRW model (orange curves) obtained from the best-fitting parameters. The shaded regions indicate the $1\sigma$ predictive uncertainties of the model.
    The right panels present the posterior probability distributions of the DRW decorrelation timescale, $\tau_{\rm DRW}$, and the asymptotic SF amplitude, $\rm SF_\infty$. The blue lines mark the median (50th percentile) values of the posterior distributions.}
    \label{fig:drwfitting}
\end{figure*}

The light curve of the DRW model is described as decorrelation function $\kappa (\Delta t) = \sigma_{\rm d}^2\, e^{-\Delta t / \tau_{\rm d}}$, where the decorrelation amplitude $\sigma_{\rm d}$ and decorrelation timescale $\tau_{\rm d}$ are the main parameters of this model. $\Delta t$ is the time lag between two measurements. 
As shown in Figure~\ref{fig:drwfitting}, J092834+292136 and J115438+065025 are well fitted with this model. Although J120052+331238 exhibits nonzero variance in all three bands as well, it cannot derive available parameters from the fitting. It is due to the weighted excess variance being sensitive to weak variability and the light-curve modeling usually requiring a higher S/N of variance.

Structure function (SF) is defined with $\sigma_{\rm d}$ and $\tau_{\rm d}$ as
\begin{equation}
    {\rm SF}(\Delta t)^2 = 2\sigma_{\rm d}^2\, (1-e^{-\Delta t/\tau_{\rm d}}).
\end{equation}
We then derive the asymptotic SF based on ${\rm SF_\infty} \approx \sqrt{2}\sigma_{\rm d}$. ${\rm SF_\infty}$ is the value SF($\Delta t \gg \tau_{\rm d}$) and represents the variability amplitude, widely used in literature. All parameters derived are listed in Table~\ref{tab:para}. 

\subsection{Estimation of Variability Upper Limit}\label{sec:upperlimit}
For all local LRDs, we use the same method to estimate the upper-limit variability amplitude as done for high-$z$ LRDs \citep{Kokubo2025ApJ...995...24K, Zhang2025ApJ...985..119Z}. \cite{Kokubo2025ApJ...995...24K} investigate the variability amplitude $\rm SF_{\infty}$ for a few LRDs based on the Bayesian method, using multiple high-sensitivity observations with JWST in at most four epochs during $\sim 2$ yr. To fully exploit observation in different bands, they consider the $\rm SF_{\infty}$ diversity along wavelength and adopt the relation of 
\begin{equation}
    \rm SF_{\infty}(\lambda) = \rm SF_{\infty}(4000\AA )\,\left (\frac{\lambda}{4000\AA }\right )^{\alpha_{var}} \label{eq:sfinfty}
\end{equation}. They then derived upper limits on $\rm SF_{\infty}(4000\AA)$ by fixing $\tau_{\rm d}$ based on the $M_{\rm BH}-\tau$ relation \citep{Burke2021Sci...373..789B} and assuming $\alpha_{\rm var} = 0 $. \cite{Zhang2025ApJ...985..119Z} extended this method to a large sample of photometry-selected LRDs and demonstrated that the majority of LRDs exhibit weak variability and only a few of them show significant variance. 

In contrast to their sample, our sample has better cadences and time baselines from the ground-based facilities. We determine the upper limit on $\rm SF_{\infty}$ using the single-band light curve with rest-frame $\lambda_{\rm eff} \approx 3800-4000 \rm \AA$, avoiding the assumption of a fixed $\alpha_{\rm var}$ and thus eliminating the uncertainty from $\alpha_{\rm var}$. We follow equation(B11) in \cite{Kokubo2025ApJ...995...24K} constructing the likelihood but for single-band difference $\Delta m_{ij}$ at a time lag of $\Delta t_{ij} = t_j-t_i$:
\begin{equation}
    \log\, p(\sigma_{d}\mid\tau_d)
        = \sum_{i<j}
        -\frac12\left[
        \frac{\Delta m_{ij}^2}{\mathrm{Var}_{ij}}
        +\log(2\pi\,\mathrm{Var}_{ij})
        \right],
\end{equation}
where variance ${\rm Var}_{ij} = {\rm SF}(\Delta t_{ij})^2\, + \sigma_{m, i}^2 \, +\sigma_{m, j}^2$ and $\sigma_{m, j}^2$ and $\sigma_{m, i}^2$ are the photometric error at moment $i$ and $j$, respectively. We derive the $\rm SF_{\infty}$ upper limit with a 90\% confidence level from posteriority. 

\begin{table*}
\centering
\caption{Variability measurement of local analogs of LRDs.} \label{tab:para}
\setlength{\tabcolsep}{2mm}{

\begin{tabular}{cccccccccc}
\hline \hline
\multirow{2}{*}{Name} & \multirow{2}{*}{R.A.}& \multirow{2}{*}{Dec.}& \multirow{2}{*}{$z_{\rm spec}$}& \multirow{2}{*}{$N_{\rm obs}$} & median $g$ & $\log M_{\rm BH}$ & $SF_{\infty}$ & $SF_{\infty, \rm DRW}$ & $\tau_{\rm DRW}$ \\
 & & & & & (mag) & ($\rm M_{\odot}$) & (mag) & (mag) & (days)\\
 (1) & (2)  & (3)  &  (4)  &  (5)  &  (6)  &  (7)  & (8) & (9) &  (10) \\
\hline
J120052+331238 & 180.2182 & 33.2107 & 0.303 & 569.0 & 20.56 & 8.41(0.03) & $<$ 0.16 &-- &-- \\ 
J092834+292136 & 142.1452 & 29.3601 &  0.293 & 336.0 & 19.47 & 7.76(0.01) & $<$ 0.19 &$0.18_{- 0.02 }^{+ 0.03 }$ &$34.1_{- 10.9 }^{+ 18.6 }$ \\ 
J115438+065025 & 178.6624 & 6.8403 &  0.269 & 158.0 & 19.59 & 7.62(0.03) & $<$ 0.16 &$0.18_{- 0.03 }^{+ 0.06 } $ &$82.2_{- 34.9 }^{+ 87.9 } $ \\ 
J003515+084859 & 8.8132 & 8.8166 &  0.242 & 66.0 & 20.69 & 6.99(0.02) & $<$ 0.02 &-- &-- \\ 
J092359+415736 & 140.9996 & 41.9602 &  0.323 & 202.0 & 21.13 & 6.86(0.04) & $<$ 0.17 &-- &-- \\ 
J122245+360218 & 185.6905 & 36.0384 &  0.301 & 205.0 & 21.28 & 6.69(0.11) & $<$ 0.13 &-- &-- \\ 
J085058+303053 & 132.7421 & 30.5149 &  0.280 & 210.0 & 21.01 & 6.29(0.18) & $<$ 0.18 &-- &-- \\ 
\hline

\end{tabular}
}
\begin{tablenotes}
    \item Note. Column (1): galaxy name. Column (2) and (3): coordinates. Column (4): redshift. Column (5): measurement number of the light curve. Colum (6): median $g$-band magnitude of the light curve. Column (7): logarithmic black hole mass, with the measurement uncertainty given in parentheses. Column (8): 90\% upper limit on the asymptotic SF, $\rm SF_\infty$. Columns (9) and (10): asymptotic SF and decorrelation timescale derived from DRW light-curve modeling.
\end{tablenotes}
\end{table*}

\begin{figure}
    \centering
    \includegraphics[width=1\linewidth]{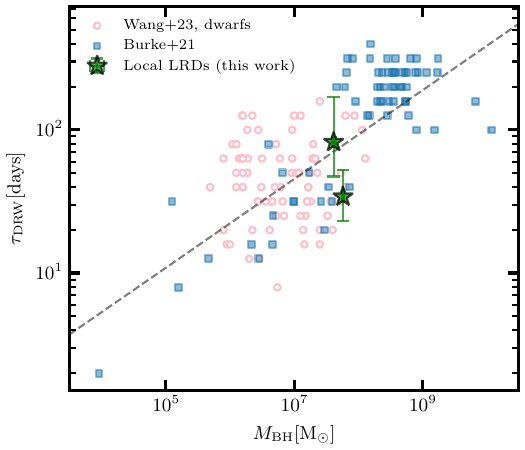}
    \caption{Decorrelation timescale $\tau_{\rm DRW}$ {\it vs.} BH mass $M_{\rm BH}$. The green stars represent J092834+292136 and J115438+065025. The pink dot is the dwarf AGNs selected based on ZTF light curves \citep{Wang2023MNRAS.521...99W}. The blue square represents the quasar sample \citep{Burke2021Sci...373..789B}.}
    \label{fig:tauMbh}
\end{figure}

\begin{figure}
    \centering
    \includegraphics[width=1\linewidth]{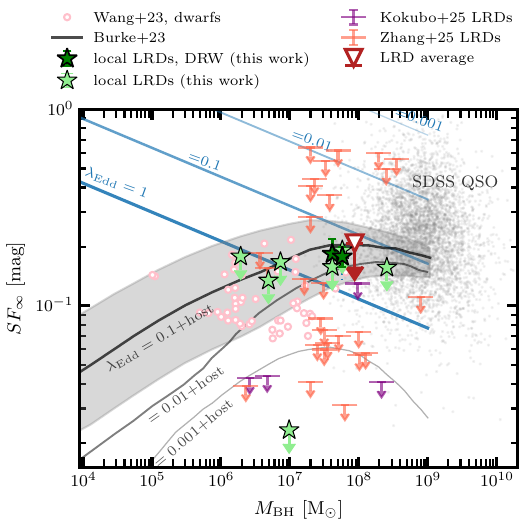}
    \caption{$\rm SF_\infty$ versus $M_{\rm BH}$. Light green stars indicate the $\rm SF_\infty$ upper limits of the seven local LRDs in our sample. Dark green stars mark the two sources (J092834+292136 and J115438+065025) for which $\rm SF_\infty$ is derived from DRW light-curve modeling. Purple and red symbols denote high-$z$ LRD samples reported by \citet{Kokubo2025ApJ...995...24K} and \citet{Zhang2025ApJ...985..119Z}, respectively, while the dark red triangle represents the average upper limit of the LRD samples. Black dots in the background show SDSS quasars taken from \cite{Burke2023MNRAS.518.1880B}. Blue curves represent the expected $\rm SF_\infty$ from AGN variability alone for different $M_{\rm BH}$ and Eddington ratios ($\lambda_{\rm Edd} \equiv L_{\rm bol}/L_{\rm Edd} = 1, 0.1, 0.01, 0.001$). Black curves show the expected $\rm SF_\infty$ when including host-galaxy dilution, assuming a color index of $g - r = 0.5$ and a covering factor of 10\% \citep{Burke2023MNRAS.518.1880B}. The gray shaded region indicates the $1\sigma$ uncertainty.}
    \label{fig:SFMbh}
\end{figure}

\section{The AGN Origin of Local LRDs}\label{sec:AGNorigin}
Figure~\ref{fig:drwfitting} presents the DRW model fits to the light curves of J092834+292136 and J115438+065025, demonstrating that their variability is consistent with AGN-like stochastic behavior. Figure~\ref{fig:tauMbh} further shows the comparison between the inferred DRW timescales $\tau_{\rm DRW}$ and black hole mass. Within the observed scatter, both sources are consistent with the empirical $M_{\rm BH}-\tau$ relation established for the AGN population \citep{Wang2023MNRAS.521...99W, Burke2021Sci...373..789B}. Their $\rm SF_\infty$ derived from the DRW model lies near the upper end of the $\rm SF_\infty$ distribution of local dwarf AGNs (see Figure~\ref{fig:SFMbh}). This does not necessarily imply that they are typical type I AGNs, as they differ from classical unobscured AGNs in both their emission-line diagnostics and SEDs.

Based on Sections~\ref{sec:drwmodel} and \ref{sec:upperlimit}, we derive the upper limits on $\rm SF_\infty$ for all sources. As shown in Figure~\ref{fig:SFMbh}, the $\rm SF_\infty$ upper limits of most local LRDs (light green stars) are clustered around 0.2 mag, lying near the upper end of the $\rm SF_\infty$ distribution observed in dwarf AGNs. Such variability amplitudes can be explained by AGN variability alone for systems accreting at sub- to super-Eddington ratios, consistent with the scenario proposed for high-$z$ LRDs. Alternatively, the observed variability may be diluted by host-galaxy light. We note that J003515+084859 exhibits an extremely low $\rm SF_\infty$ upper limit, which is most likely attributable to the low cadence of its light curves, leading to an underestimation.

Compared to high-$z$ LRDs, the distribution of $\rm SF_\infty$ upper limits for the local analogs appears more concentrated. It is worth noting that a subset of high-$z$ LRDs exhibits extremely low variability upper limits. However, the high-$z$ samples currently suffer from low cadence and short baseline, and their reported upper limits may not be as conservative as previously expected (see the discussion in Section~\ref{sec:underestimation}). Instead, their variability amplitudes may be underestimated. 
Moreover, most existing high-$z$ samples are selected photometrically \citep[e.g., ][]{Zhang2025ApJ...985..119Z} and lack spectroscopic confirmation, which may induce contamination. 

Taking into account the potential underestimation/overestimation of variability in LRD measurements, we compute the average value of $\mathrm{SF}_\infty$ upper limit for the LRD population (red triangle). This value is very close to the upper limits of the majority of local LRDs, further supporting the similarity in variability properties between local analogs and LRDs.

It is also useful to compare our results with previous variability studies of local LRDs. \cite{Burke2025arXiv251116082B} investigated the optical variability of three local LRDs reported in \cite{Lin2025arXiv250710659L} using light curves spanning $\sim 5$ yrs and reported weak {\it intrinsic variability}. In contrast to our approach, they infer the intrinsic variability by modeling light curves that have relatively large photometric uncertainties. As their sources have $g$-band magnitudes comparable to ours, the S/N of the light curves is similarly limited. In such a low-S/N regime, intrinsic variability estimates can be sensitive to the treatment of photometric uncertainties and assumed variability models. Therefore, differences in methodology may contribute to the discrepancy between the inferred variability amplitudes.

\begin{figure*}
    \centering
    \includegraphics[width=0.98\linewidth]{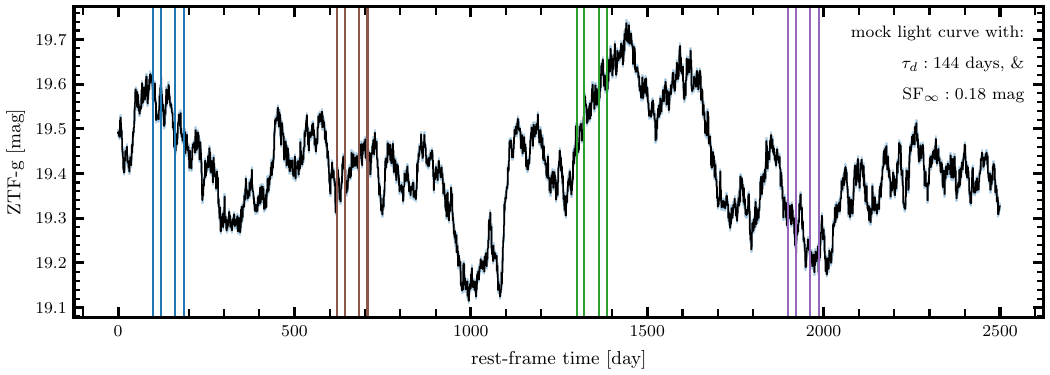}
    \includegraphics[width=0.24\linewidth]{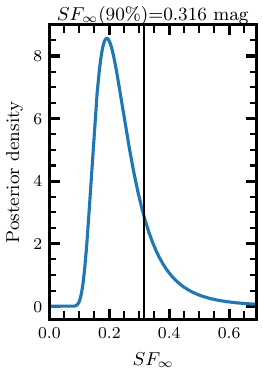}
    \includegraphics[width=0.24\linewidth]{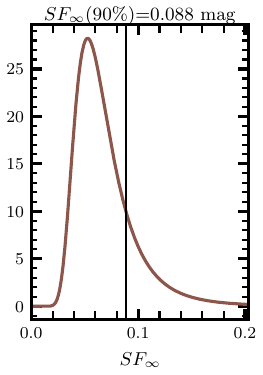}
    \includegraphics[width=0.24\linewidth]{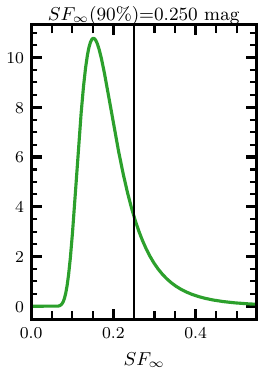}
    \includegraphics[width=0.24\linewidth]{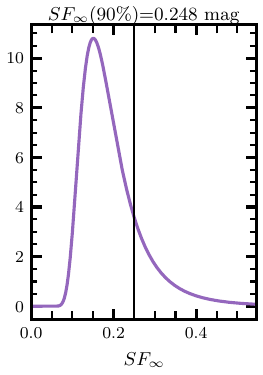}
    \caption{
    Examples of mock observations with temporal offsets of 100 (blue), 620 (brown), 1300 (green), and 1900 (purple) days.
    {\it Top}: The black curve shows a DRW light curve with $\tau_{\rm DRW}$ of 144 days and $\rm SF_\infty$ of 0.18 mag. For each offset, the rest-frame sampling epochs (0, 22, 63, and 87 days) are adopted from those of the LRD MSAID38108 \citep{Kokubo2025ApJ...995...24K}.
    {\it Bottom}: The posterior distributions of $\rm SF_\infty$. The black lines mark the 90th-percentile upper limits.
    }
    \label{fig:simulations}
\end{figure*}

\begin{figure}
    \centering
    \includegraphics[width=1\linewidth]{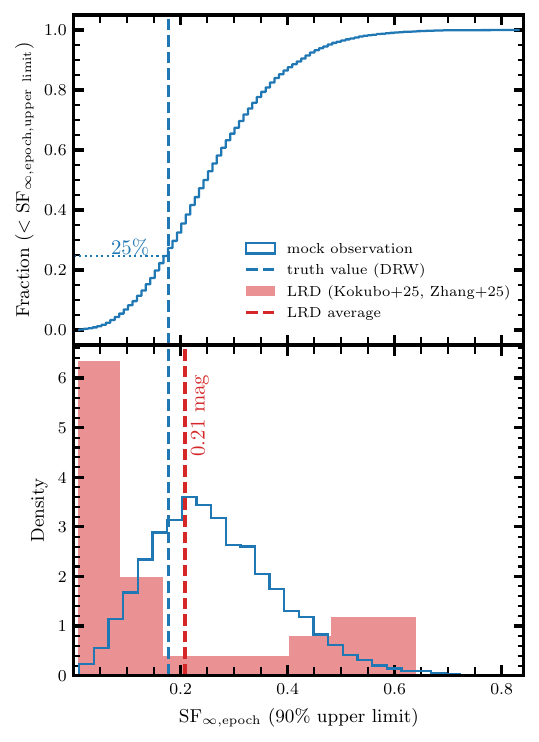}
    \caption{Accumulation fraction ({\it top}) and histogram ({\it bottom}) of the $\rm SF_\infty$ upper limits estimated from mock observations based on observations of MASID38108. The red line marks the median (50th percentile). The blue dashed lines indicate the true value of $\rm SF_\infty$ used in simulation, and the red dashed line is the average of $\rm SF_\infty$ upper limits for LRDs.}
    \label{fig:hist}
\end{figure}

\section{Current Underestimation of Variability for High-$z$ LRDs}\label{sec:underestimation}
\subsection{Mock Observation}
To better quantify the methodological uncertainties in estimating variability amplitude upper limits for high-$z$ LRDs, we perform mock observations assuming an $\rm SF_\infty$ equal to that of J092834+292136, under the assumption that high-$z$ LRDs exhibit variability amplitudes comparable to local analogs.

We first construct 50 realizations of high-cadence model light curves spanning 2500 days based on the DRW model. We adopt the solution given in Equation (10) of \cite{Moreno2019PASP..131f3001M}, assuming an $\rm SF_\infty$ of 0.18 mag (equal to that of J092834+292136) and an intrinsic timescale $\tau_{\rm int}$ of 144 days, derived from the $M_{\rm BH}-\tau$ relation using the black hole mass of LRD MSAID38108 \citep{Kokubo2025ApJ...995...24K}.

For each realization, we evaluate the model fluxes at the rest-frame epochs (0, 22, 62, and 87 days), corresponding to the cadence of the JWST epoch pairs for MSAID38108. This procedure generates a set of sampled light curves that mimic the JWST LRD observations. 

We measure the $\rm SF_\infty$ upper limit with a method cosistent to the one in Section \ref{sec:upperlimit}. 
Specifically, we adopt 144 days for MSAID38108 with $M_{\rm BH} = 10^{8.34}\, M_\odot$, which matches the intrinsic timescale of the simulated light curves. Four illustrative examples are shown in Figure~\ref{fig:simulations}.

Because AGN variability is stochastic, the observed variability amplitude can increase when the observing epochs coincide with a flaring segment. To account for this effect, we repeat the above procedure while shifting the epoch set forward by 10 days at each iteration. Over the $\sim2500$ day rest-frame baseline of the mock light curve, this yields 240 independent estimates of the upper limit on $\mathrm{SF}_\infty$ per realization, resulting in 12,000 measurements in total across all 50 realizations.

Although we adopt a specific epoch set corresponding to a particular LRD, this source has the longest baseline in the sample of \cite{Kokubo2025ApJ...995...24K}. We verify that using epoch sets drawn from other LRDs with shorter baselines yields nearly identical results (see Appendix~\ref{sec:appendix} for details). This indicates that the dominant uncertainty arises from a short baseline of currently limited observation, rather than from the details of any individual object.

\subsection{Causes of the Underestimation}\label{sec:underestimation}
Figure~\ref{fig:hist} shows the distribution of the $\rm SF_{\infty}$ upper limits derived from 12,000 mock observations. We find that in approximately 25\% of the cases, the inferred upper limits (90\% confidence level) are lower than the input value of $0.18 \rm\, mag$, which corresponds to the true $\rm SF_{\infty}$ obtained from the DRW fit.

For a properly derived 90\% upper limit, only $\sim$10\% of observations are expected to yield upper limits smaller than the true value. The higher fraction found in our simulations, therefore, indicates a systematic underestimation of the upper limits of variability amplitude under the current observational conditions.

This effect appears frequently in our simulations and likely explains why previous studies \citep[e.g.,][]{Kokubo2025ApJ...995...24K, Zhang2025ApJ...985..119Z, Tee2025ApJ...983L..26T} reported extremely weak variability amplitudes for high-$z$ LRDs. When treating the LRD sample as a whole, the average $\rm SF_{\infty}$ upper limit of high-$z$ LRDs is $0.21\, \rm mag$ (red triangle in Figure~\ref{fig:SFMbh}), which is comparable to that measured for local analogs.

This underestimation likely arises from short temporal baseline relative to the intrinsic timescale. Our simulations show that when the ratio between the light-curve baseline and the intrinsic DRW timescale is sufficiently large (baseline/$\tau_{\rm int} \sim 2$), the 90\% upper limits are correctly recovered in more than 90\% of realizations (see Appendix~\ref{sec:appendixB} in details). However, when the baseline becomes comparable to or shorter than the intrinsic timescale, the failure rate increases. In \cite{Kokubo2025ApJ...995...24K}, the adopted variability timescales for several LRDs are comparable to or even longer than the observational baselines. Under such conditions, the posterior distribution of $\rm SF_{\infty}$ becomes prior driven and tends to bias the upper limits toward lower values.

An alternative factor can be the scatter in the empirical $M_{\rm BH}-\tau$ relation. The adopted decorrelation timescales are typically inferred from empirical $M_{\rm BH}-\tau$ relations calibrated for the general AGN population. However, these relations exhibit substantial intrinsic scatter of $0.46^{+0.09}_{-0.10}$ dex \citep{Wang2023MNRAS.521...99W}. In addition, given the uncertainty in the physical origin of broad lines in LRDs \citep{Naidu2025arXiv250316596N, Rusakov2026Natur.649..574R, Zhang2026NatAs.tmp...41Z, Juodzbalis2025arXiv250821748J}, the black hole masses derived from virial estimates may also be biased. If the intrinsic variability timescale of an individual LRD differs significantly from the assumed value, the inferred $\rm SF_{\infty}$ upper limit can be systematically biased.

Given the limited temporal coverage of current JWST observations, the balance between the intrinsic variability timescale and the observational baseline appears to be a key factor governing the robustness of the derived upper limits.

\section{Conclusions}\label{sec:conclusion}
Based on ZTF light curves spanning about 6 yr, we investigate the optical variability of seven local analogs of LRDs. We estimate the upper limits of variability amplitude using the same method applied to high-$z$ LRDs. 

Among these local analogs, the three brightest sources show variances in all three ZTF bands. Two of them are well fitted by a DRW model, supporting an AGN origin for their variability. The decorrelation timescales derived from the DRW model are also consistent with those of dwarf AGNs and quasars.

Compared to high-$z$ LRDs, local analogs have light curves with longer temporal baselines, and the resulting distribution of variability amplitude upper limits is narrower and close to the average value of upper limits for high-$z$ LRDs. This further supports the similarity between LRDs and their local analogs in terms of variability.

To better understand the limitations of the current methodology for measuring variability in high-$z$ LRDs given limited observations, we generate mock observations with the same temporal cadence as high-$z$ LRDs. We find that their variability amplitude upper limit is probably underestimated, especially when the temporal baseline is shorter than the intrinsic timescale. This suggests that the dominant uncertainty in current LRD variability constraints arises from the short baselines of the available light curves derived from JWST, as well as the large intrinsic scatter in the empirical $M_{\rm BH}-\tau$ relation.

\begin{acknowledgments}
This work is supported by the National Key R\&D Program of China No.2022YFF0503402. We also acknowledge the science research grants from the China Manned Space Project, especially, NO.  CMS-CSST-2025-A18. ZYZ acknowledges the support of the Shanghai Leading Talent Program of Eastern Talent Plan (LJ2025051) and the China-Chile Joint Research Fund (CCJRF No. 1906). JXW was supported by the National Science Foundation of China (12533006). LCH was supported by the National Science Foundation of China (12233001) and the China Manned Space Program (CMS-CSST-2025-A09).

Based on observations obtained with the Samuel Oschin Telescope 48-inch and the 60-inch Telescope at the Palomar Observatory as part of the Zwicky Transient Facility project. ZTF is supported by the National Science Foundation under Grants No. AST-1440341 and AST-2034437 and a collaboration including current partners Caltech, IPAC, the Oskar Klein Center at Stockholm University, the University of Maryland, University of California, Berkeley, the University of Wisconsin at Milwaukee, University of Warwick, Ruhr University, Cornell University, Northwestern University and Drexel University. Operations are conducted by COO, IPAC, and UW.

Some of the data products presented herein were retrieved from the Dawn JWST Archive (DJA). DJA is an initiative of the Cosmic Dawn Center (DAWN), which is funded by the Danish National Research Foundation under grant DNRF140.

\end{acknowledgments}

\begin{contribution}

RQL led the project and performed the data analysis. ZYZ contributed to the original concept and helped define the overall scientific framework. All authors discussed the results and provided comments to the manuscript.


\end{contribution}

%
\facilities{ZTF \citep{Masci2019PASP..131a8003M}}

\software{ZTF Lightcurves \citep{https://doi.org/10.26131/irsa598}, {\tt astropy} \citep{2013A&A...558A..33A,2018AJ....156..123A,2022ApJ...935..167A},
        \texttt{celerite} \citep{celerite1, celerite2},
        \texttt{emcee} \citep{Foreman-Mackey2013PASP..125..306F},
          {\tt grizli} \citep{brammer_2023_8370018},
          {\tt msaexp} \citep{brammer_2023_8319596},
          }


\appendix

\section{Mock observation using cadence of various LRDs }\label{sec:appendix}
\begin{figure}
    \centering
    \includegraphics[width=0.5\linewidth]{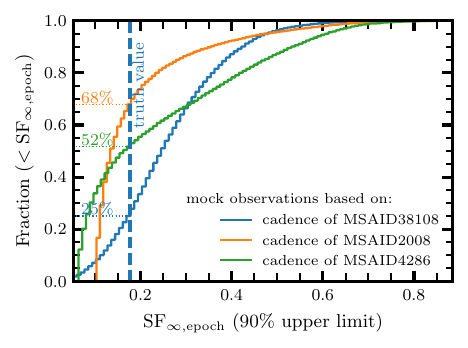}
    \caption{Accumulation fraction of $\rm SF_\infty$ upper limits estimated from mock observations based on observations of MASID38108 (blue), MASID2008 (orange), and MASID4286 (green). The blue dashed lines indicate the true value of $\rm SF_\infty$ used in simulation.}
    \label{fig:mulriCadence}
\end{figure}

Here, we simulate observations using corresponding cadences and $\tau_{\rm int}$ inferred from black hole mass for individual LRDs and derive the accumulation fraction of $\rm SF_\infty$ upper limits. We adopt the observing cadences of three LRDs, MASID38108, MASID2008, and MASID4286 \citep{Kokubo2025ApJ...995...24K}, whose rest-frame cadences are (0, 2, 62, 87), (0, 34, 45), and (0, 6, 54) days, respectively. As shown in Figure~\ref{fig:mulriCadence}, for the cadences of MASID38108, MASID2008, and MASID4286, 25\%, 52\%, and 68\% of the mock observations yield $\rm SF_\infty$ upper limits lower than the true value, respectively.

\section{Mock observation using different intrinsic timescales} \label{sec:appendixB}
\begin{figure}
    \centering
    \includegraphics[width=0.5\linewidth]{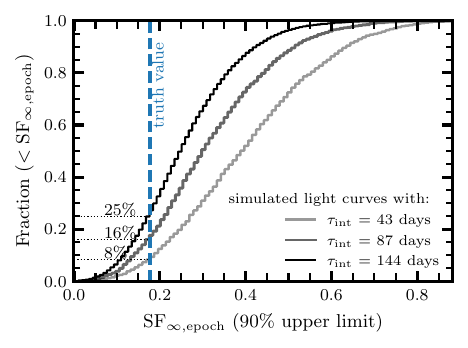}
    \caption{Accumulation fraction of $\rm SF_\infty$ upper limits assuming various intrinsic timescales of realization. The vertical blue lines indicate the true value of $\rm SF_\infty$ used in simulation, and the red line is the average of $\rm SF_\infty$ upper limits for LRDs.}
    \label{fig:mulriCadence2}
\end{figure}

In addition to the simulation described in Section~\ref{sec:underestimation}, we also generate realizations with shorter intrinsic timescales, $\tau_{\rm int}$ of 300, 87, and 43 days, corresponding to 2, 1, and 1/2 times the rest-frame baseline, respectively. We retain the assumption of $\tau_{d} = 144$ days when deriving the $\rm SF_\infty$ upper limits. As shown in Figure~\ref{fig:mulriCadence2}, the success rate of correctly recovering the upper limits increases as $\tau_{\rm int}$/$\tau_{d}$ decreases. 

\bibliography{ref}{}
\bibliographystyle{aasjournalv7}


\end{CJK*}
\end{document}